\documentstyle[sprocl,epsf]{article}
\def\beq{\begin{equation}}
\def\eeq{\end{equation}}
\def\beeq{\begin{eqnarray}}
\def\beeqn{\begin{eqnarray*}}
\def\eeeq{\end{eqnarray}}
\def\eeeqn{\end{eqnarray*}}
\def\ie{\hbox{\it i.e.}{ }}      
\def\etc{\hbox{\it etc.}{ }}
\def\eg{\hbox{\it e.g.}{ }}      

\def\aka{\hbox{\it a.k.a.}{ }}
\def\half{\mbox{\small $\frac{1}{2}$}}

\def\nonp{non-perturbative}
\def\phi{\varphi}

\begin{document} 
\title{PROPERTIES OF DERIVATIVE EXPANSION APPROXIMATIONS TO THE 
RENORMALIZATION GROUP}
\author{ T.R. MORRIS }
\address{Department of Physics, University of Southampton, Highfield,\\
Southampton SO17 1BJ, UK}
\maketitle\abstracts{ 
Approximation only by derivative (or more generally momentum) expansions,
combined with reparametrization invariance, turns the continuous 
renormalization group 
into a set of partial differential equations
which at fixed points become non-linear eigenvalue equations for the
anomalous scaling dimension $\eta$. We review how these equations provide a
powerful and robust means of discovering and approximating non-perturbative
continuum limits. Gauge fields are briefly discussed. Particular emphasis
is placed on the r\^ole of reparametrization invariance, and the 
convergence of the derivative expansion is addressed.}

This talk is about derivative (or more general momentum) expansions as
applied to the renormalization group, in a quantum field theory setting.
The motivation is simply this: I want to construct analytic approximation
methods with as much reliability and accuracy as possible {\sl even when
there are no obviously small parameters}, \eg $\epsilon=4-D$, $1/N$ \etc
with which one could expand perturbatively.
(Here, $D$ is space-time dimension, and $N$ is the number of components of 
the field.) In other words, I want to look for approximations that work 
in a genuinely \nonp\ setting. Now, as Wilson was instrumental in
demonstrating, the continuum limit of a quantum field theory can \nonp ly
be best understood in terms of the flow of an effective action 
$S_\Lambda[\phi]$ under lowering an effective U.V. (ultra-violet) 
cutoff $\Lambda$.\cite{kogwil} 
Thus in this framework one works with a flow equation
that generically takes the form
\beq
\Lambda{\partial\over\partial\Lambda} S_\Lambda[\phi] 
= {\cal F}[S_\Lambda]\quad,
\eeq
where the cutoff is implemented through some 
effective U.V. cutoff function $C_{UV}(q,\Lambda)$. (Here $q$ stands
for momentum, and the above equation is referred to as
the continuous, or momentum space,
renormalization group.) Scale invariant continuum limits
(thus massless field theories) are then simply given by 
fixed points: \footnote{once all quantities have been
rewritten in terms
of dimensionless quantities, using $\Lambda$}
\beq
\Lambda{\partial\over\partial\Lambda} S_\Lambda[\phi] =0\quad.
\eeq
 The massive continuum
limits follow from  
tuning the relevant perturbations around
these fixed points. In such a setting one realises that various
approximations can be made quite easily that preserve the structure
of the continuum limit, while in other frameworks (for example when
using truncations of Dyson-Schwinger equations) the continuum limit,
equivalently renormalisability, is almost inevitably destroyed.\cite{erg}

What are the possible approximations? The first thought is to
try truncating the space of interactions to just a few operators,
however this results in a truncated expansion 
 in powers of the field $\phi$ (about some point).
Such an approximation can only be sensible if the field $\phi$ does
not fluctuate very much, which is the same as saying
 that it is close to mean field,
\ie in a setting in which weak coupling perturbation theory is anyway valid.
Studying the behaviour of truncations
in a truly \nonp\ situation, one finds that higher orders cease
to converge and thus yield limited accuracy, while there is also
no reliability -- even qualitatively -- since many spurious
fixed points are generated.\cite{trunc}

This situation should be contrasted with truncations to a few operators
in the real space renormalization group of spin systems, such as block
spin renormalization group of the Ising model.
Such truncations were extensively studied in the past,\cite{rsRG} and could
be very accurate.\cite{burk} A modern variant produces 
spectacularly accurate results in low dimensions.\cite{white} The powerful
Monte-Carlo renormalization group
methods,\cite{MCrev}
 are also based on such truncations.
 In the case of such simple discrete systems however, the expansion
effectively results in a short distance expansion of the effective action,
for example by keeping only the finite number of
interactions linking nearest neighbours,
then next-to-nearest neighbours, and so on. 

The analogous expansion in our continuum case is, for smooth cutoff
functions $C_{UV}$, a derivative expansion of $S_\Lambda[\phi]$ (equivalent
to a Taylor expansion in the momenta of its vertices),\cite{deriv,twod}
while for
sharp cutoff functions $C_{UV}(q,\Lambda)=\theta(\Lambda-q)$ it is an
expansion in momentum scale $\sim |\partial/\Lambda| \equiv p/\Lambda$,
where the coefficients  are not analytic in $p^\mu$ but rather, 
non-trivial functions of the angles between various momenta which
must be determined self-consistently through the flow equation.\cite{truncm}
[The non-analyticity is a purely technical problem that is induced by the
non-analyticity of $\theta(\Lambda-q)$.] At any rate, in both cases
such a short distance expansion -- {\sl where no other approximation
is made} -- seems a particularly natural approximation to try, and in view
of the discussion above, sensible results might well be expected providing
the Wilson effective action $S_\Lambda$ is `sufficiently well behaved':
thus the approximation would fail if the higher derivative 
terms are not in some sense small, but this would indicate that a description
in terms of the given field content is probably itself inappropriate
and other degrees of freedom should be introduced. This is an important
point, to which I will return later.

Consider the case of $O(N)$ invariant scalar field theory,
the so-called $N$-vector model. I shall start by 
using  a sharp cutoff  
and making the simplest such approximation --
keeping only a potential interaction:
\beq
S_\Lambda\sim\int\!\!d^D\!x\,\left\{\half(\partial_\mu\phi^a)^2+V(\phi,\Lambda)
\right\}\quad.
\eeq
After appropriate scalings to dimensionless combinations, the flow equation
is found to be:
\beq
\label{flo}
-\Lambda{\partial\over\partial\Lambda}V+d\,\phi V'
-DV=\ln(1+V'')+(N-1)\ln(1+{V'/\phi})\quad.
\eeq
Here $\phi\equiv\sqrt{\phi^a\phi^a}$, $\prime\equiv\partial/\partial\phi$,
and I have introduced the dimension of the field $\phi$: $d={D/2}-1$.
(Since we have thrown away all momentum dependent corrections, $\eta=0$
in this approximation.)
The $N=\infty$ case of this equation was already derived by Wegner and
Houghton in their paper introducing the sharp-cutoff flow equation,\cite{wegho}
and subsequently the general $N$ case was proposed as a ``local potential''
approximation by Nicoll, Chang and Stanley.\cite{nico} It has since
been rediscovered by many authors,\cite{truncm} especially Hasenfratz and 
Hasenfratz.\cite{hashas} Nevertheless this equation, and its
smooth cutoff sisters, have a number of beautiful
properties that have not been pointed out by previous workers.

First note that the fixed point equation
for $V(\phi,\Lambda)\equiv V(\phi)$,
\beq
\label{fpe}
d\,\phi V'
-DV=\ln(1+V'')+(N-1)\ln(1+{V'/\phi})\quad,
\eeq
determines {\sl by itself} at most a countable 
set \footnote{$D=2$ is an exception: see later} of 
sensible fixed point potentials,
each of which can be identified with approximations to the exact fixed points.
This is not obvious because eqn.(\ref{fpe}) is a second order ODE
(Ordinary Differential Equation) 
and therefore in some neighbourhood of some starting value $\phi$ we
can construct a continuously infinite two parameter set of solutions.
Actually, all but a countable number of those solutions are singular!
For example,  consider the case $D=4$ and $N=4$ (the
Higgs field in the
Standard Model). Obviously from (\ref{fpe}),
 $V'(0)=0$ is necessary if the potential is not to be
singular at $\phi=0$. We can choose a value of $V(0)$ and then 
(numerically) integrate
out to $\phi>0$ using eqn.(\ref{fpe}). One discovers almost always
that at some point
$\phi=\phi_c$ a singularity in $V$ is encountered, after which the
potential ceases exist (or at least is complex for $\phi>\phi_c$).
The first graph in fig.1 is a plot of 
$\phi_c$ against $V(0)$. We see that only
the (trivial) Gaussian fixed point solution $V(\phi)\equiv0$ exists for 
all values
of the field. If the same is done for the case $D=3$ and $N=1$, we get the
second graph in fig.1. In this case there is also one non-trivial non-singular
solution, corresponding to the famous Wilson-Fisher fixed point
(Ising model universality class). I have
checked all cases $D=3,4$ and $N=1,2,3,4$;  they reproduce the
standard fixed points.\cite{trunc} 

\vskip -2cm
\begin{figure}[ht]  
\epsfxsize=0.9\hsize
\epsfbox{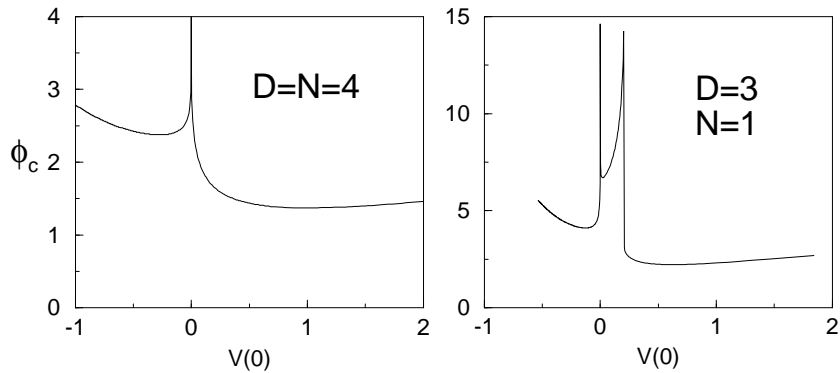} 
\vskip -2cm
\caption{Plots of $\phi_c$ against $V(0)$ for $D=N=4$, and $D=3$ and
$N=1$.}
\end{figure}

Studying eqn.(\ref{fpe}),
one can convince oneself that the only way
that $V$ can satisfy this equation as $\phi\to\infty$ is if 
$V(\phi)\sim\phi^{D/d}$. Together with $V'(0)=0$,  
we now have two boundary conditions and thus we should expect only
a countable number of solutions from the second order ODE.
For $D=2$, exceptions arise (due to $d=0$), thus for $N=1$
one obtains a semi-infinite continuous line of fixed points with
periodic potentials, but these may be identified with critical sine-Gordon
models.\cite{twod} On the other hand, there is no need to impose
$V'(0)=0$ when $N=1$. This just allows a constant 
phase shift on critical sine-Gordon potentials, 
while in higher than two dimensions the power law constraint on
$V$ now holds separately for $\phi\to\infty$ and $\phi\to-\infty$ (\ie
with possibly
different coefficients). Nevertheless, 
we have confirmed that
 in this larger space, there is
still only the one 
non-trivial fixed point in three dimensions. 

 While all this just reproduces the standard lore, note
nevertheless how powerful the method is: the {\sl infinite dimensional}
space of {\sl all possible potentials} $V(\phi)$ has been searched for 
continuum limits. Clearly this is much more than is possible with other
methods! Also, the continuum is actually accessed {\sl directly} without the
need to go through the construction of introducing
 an overall cutoff $\Lambda_0$, a bare action $S_{\Lambda_0}$,
 and then taking the continuum limit
$\Lambda_0\to\infty$. 

These properties are true also when the approximation is applied to
the massive theory. In this case one must determine the form of the
perturbations about the fixed point. One can write
\beq
V(\phi,\Lambda)=V(\phi)+\epsilon\, v_\lambda(\phi) \,\Lambda^{-\lambda}\quad,
\eeq
where eqn.(\ref{flo}) is linearised in $\epsilon$ and separation of
variables has been used. Now, $v_\lambda(\phi)$ satisfies a linear
second order ODE, and once again we
appear to have a continuously infinite set of solutions, and
for all choices of $\lambda$. In this case the crucial
observation is that if, beyond the linearised level,
 the scale dependence of the perturbation is to be absorbed into an associated
coupling $g_\lambda(\Lambda)$, \ie a renormalised coupling corresponding to
universal self-similar flow about the fixed point, then $v_\lambda(\phi)$
has to behave as $v_\lambda(\phi)\sim \phi^{(D-\lambda)/d}$
as $\phi\to\infty$. \cite{hh}
For the same reasons as before, this typically allows 
only a countable number of solutions, but 
this time we also have linearity, which
 implies a normalization condition can be set,
overconstraining the equations and resulting in quantization of $\lambda$.
(Again, $d=0$ provides an exception -- resulting in
 more general perturbations with \eg exponential or periodic behaviour. 
It is worth remarking that for $d\ne0$, the power laws given above
are the unique powers required so that
the physical \footnote{\ie in the original dimensionful variables.} 
 $v_\lambda$ and $V$  are independent of $\Lambda$,  and therefore 
obtain a non-trivial finite limit as $\Lambda\to0$; this limit gives  
the Legendre effective potential \cite{erg} and thus the 
equation of state.)

Consider now the derivative
expansion at $O(\partial^2)$. In this case we need to use a smooth cutoff
(as already discussed). The effective action takes the form
\beq
S_\Lambda\sim\int\!\!d^D\!x\,\left\{V(\phi)+
\half(\partial_\mu\phi^a)^2 K(\phi)+
\half(\phi^a\partial_\mu\phi^a)^2 Z(\phi)\right\}\quad
\eeq
(where this last term is required only for $N\ne1$).
In this case the fixed point equations are a set of coupled 
second-order non-linear ODEs, one for each coefficient function ($V$, $K$
and $Z$). This pattern holds to all orders of the derivative
expansion. As previously, one can argue for specific power law behaviours
for $V$, $K$ and $Z$, and that typically only a countable number of non-singular
solutions exist. But now there is another parameter to determine: the
critical exponent $\eta$ from the anomalous scaling of $\phi$. The
{\sl exact} renormalization group has a $\phi$ reparametrization 
invariance \cite{red,rie} reflecting the fact that physics is independent 
of the normalization of the field, and this extra invariance turns the fixed
point equations into non-linear eigenvalue equations for $\eta$,
because it allows a normalization condition (\eg $K(0)=1$)
and thus quantization of $\eta$
(in a similar way to the linear case above for perturbations.)
There is a problem however: the derivative expansion generally breaks
the reparametrization invariance, with the result that $\eta$, $\nu$,
$\omega$ \etc depend on some unphysical parameter such as $K(0)$, the
normalization of the kinetic term. \cite{gol}

Consider the Polchinski form \cite{pol} 
of the Wilson flow equation. Schematically, 
\beq
\label{Pol}
{\partial S_\Lambda \over\partial\Lambda}={1\over2}\,{\rm tr}\, 
{\partial\Delta_{UV}\over\partial\Lambda}
\left\{ {\delta S_\Lambda\over\delta\phi}
{\delta S_\Lambda\over\delta\phi}-{\delta^2S_\Lambda\over\delta\phi\delta\phi}
-2\left(\Delta^{-1}_{UV}\phi\right)
{\delta S_\Lambda\over\delta\phi}\right\}\quad,
\eeq
where $\Delta_{UV}(q,\Lambda)=C_{UV}/q^2$, and 
the total action $S_\Lambda \sim{1\over2}\,\phi.\Delta_{UV}^{-1}.\phi\,
 + S^{int}_\Lambda$.
This equation is simply related to the Wilson
equation \cite{kogwil}
 through $\phi\mapsto\sqrt{C_{UV}}\,\phi$ 
and ${\cal H}\equiv-S_\Lambda^{int}$, \cite{deriv,prep}
but in contrast to Wilsons, it has
 the intuitively nice property that eigen perturbations
about the Gaussian fixed point are precisely polynomials in the field and its
derivatives. Now in general, the reparametrization 
symmetry is given by a complicated functional integral transform,\cite{rie}
so it is not surprising that a truncated
derivative expansion destroys it --
and in fact it is far from clear how to approximate at all
in a way which preserves it. 

(Let me 
 emphasise that with broken reparametrization invariance,
defining $z(\phi)$ through
$S^{int}_\Lambda\sim\int\ V(\phi)+\half z(\phi)(\partial_\mu\phi^a)^2 +\cdots$,
the results  {\sl depend} on the
 value $z(0)$. This dependence on $z(0)$ has
not been recognized by authors who
set $z(0)=0$ with insufficient justification.\cite{ball})

Fortunately, for two special forms of cutoff function,
reparametrization invariance may be linearly realized.
These correspond to either sharp cutoff
$C_{UV}=\theta(\Lambda-q)$ or a power law smooth cutoff 
$C_{UV}\sim 1/[1+(q/\Lambda)^{2\kappa+2}]$, where
$\kappa$ is some non-negative integer.
(The underlying reason is that such cutoffs are left invariant
by a subgroup of linearly realised `universality symmetries' that
map between different schemes in (\ref{Pol}),\cite{prep} but it would take us
too far afield to explain this.) 
However, a direct derivative/momentum-scale
expansion of the Wilson flow
equation leads to singular coefficients with both of these 
cutoffs.\cite{erg,twod,truncm,ball} The way out of this difficulty is to
 recognize that, from the
form of the right hand side of eqn.(\ref{Pol}), the Wilson effective
action has a tree structure.\cite{pol}
It is the Taylor expansion of the corresponding propagators  that
causes the problem.\cite{erg,twod,truncm,ball}
Therefore, to overcome this difficulty we first pull
out the one particle irreducible parts.

It can be shown \cite{erg} that the one particle irreducible parts
of $S_\Lambda$ are
generated by a Legendre effective action $\Gamma_\Lambda[\phi]$
equipped with infrared cutoff
$C_{IR}=1-C_{UV}$, \ie related in the usual way to a partition function
except that the bare action is modified to
$S_{\Lambda_0}=\half\phi.\Delta_{IR}^{-1}.\phi+ S^{int}_{\Lambda_0}$,
where $\Delta_{IR}=C_{IR}/q^2$. Whence, 
the `Legendre flow equation'  follows; \cite{nici,erg,wet} 
\beq
{\partial\Gamma^{int}_\Lambda\over\partial\Lambda} = -{1\over2}\, {\rm tr}\,
\left[ {1\over\Delta_{IR}}
{\partial\Delta_{IR}\over \partial\Lambda} . \left( 1+ \Delta_{IR}.
{\delta^2\Gamma^{int}_\Lambda\over\delta\phi\delta\phi}
\right)^{-1}\right]\quad.
\eeq
In this form, the sharp cutoff limit 
enjoys the simplest reparametrization invariance:\cite{truncm}
$\phi\mapsto a\,\phi$. 
For the smooth power law cutoff, certain momentum independent
linear transformations on other quantities
are also required.\cite{deriv} 
Since these transformations are linear and momentum independent, a
truncated derivative expansion now {\sl preserves} the reparametrization
invariance, while derivative/momentum-scale
 expansion of $\Gamma^{int}_\Lambda$ is well-defined
with these cutoffs,\cite{deriv,truncm}
because it results in Taylor expansion of the self-energy,
rather than the propagator itself.

Returning now to $O(\partial^2)$, we need the smooth cutoff 
and  choose the integer $\kappa$ as small as possible, to
maximise the accuracy of the derivative expansion.\cite{deriv}
Table 1 displays the results obtained in
$D=3$ dimensions,\cite{deriv,oN} for $\eta$ and $\nu$,\footnote{from which 
all other exponents (excepting
correction to scaling exponents) follow, because all hyper-scaling 
relations are here exactly preserved} and for the first correction to scaling
exponent $\omega$. Only the expected fixed points are found.
\begin{table} [ht]
\renewcommand{\arraystretch}{1.5}
\hspace*{\fill}
\begin{tabular}{|c||c|c||c|c|c||c|c|c|}  \hline
$N$ &\multicolumn{2}{c||}{$\eta$} &\multicolumn{3}{c||}{$\nu$}
&\multicolumn{3}{c|}{$\omega$} \\ \hline & $O(\partial^2)$ & World &
$O(\partial^0)$ & $O(\partial^2)$ & World & $O(\partial^0)$ &
$O(\partial^2)$ & World 
\\ \hline 1 & .054 &.035(3) &.66 &.618 &.631(2) &.63 &.897 &.80(4) 
\\ \hline 2 & .044 &.037(4) & .73 & .65 & .671(5) &.66 & .38&.79(4) 
\\ \hline 3 & .035 &.037(4) & .78 & .745 &.707(5)&.71 & .33&.78(3)
\\ \hline 4 & .022 &.025(4) & .824 & .816 &.75(1) & .75 & .42  & ? 
\\ \hline\hline
          10& .0054 &.025 & .94 & .95 & .88& .89 & .82 & .78 
\\ \hline 20& .0021 & .013& .96 & .98 & .94& .95 & .93 & .89 
\\ \hline 100 & .00034 &.003 & .994& .998 &.989 & .991 & .988 & .98 \\
\hline
\end{tabular}
\hspace*{\fill}
\renewcommand{\arraystretch}{1}
{\caption[Critical exponents of the 
three-dimensional Wilson-Fisher fixed point.]
{ Critical exponents of the three-dimensional Wilson-Fisher fixed point.
The first two orders of the derivative expansion are compared to
a combination of the worlds best estimates,\cite{zinn,etcN,N4} with 
their errors, where available. $\eta$ is identically zero for all $N$ at
$O(\partial^0)$.}}
\label{t:2crit}
\end{table}
For $N=1\cdots4$, the results are fair
for the local potential approximation and already quite good 
at $O(\partial^2)$, with the exception of $\omega$ which gets worse at $O(\partial^2)$ when $N>1$. 
For the large $N$ cases however, $\omega$
is better estimated, $\nu$ is not improved at $O(\partial^2)$,
while $\eta$ is dramatically underestimated --
eventually by about a factor of 10. 
Something is going wrong particularly at
large $N$. Actually, the results for $N=\infty$ are guaranteed
to be right because in this case the $O(\partial^0)$ approximation is
exact,\cite{wegho,oN} so the problem only appears in the approach to this
limit. I believe that this is an example of a case where all the appropriate
fields have {\sl not} been included in the effective action. Indeed,
 it is known that at large $N$, a massless bound state
field also exists at the critical point.\cite{zinn,etcN} We should thus
expect that a derivative expansion is ill-behaved, because the vertex
functions are hiding within them the effects of this integrated out 
massless field. To ameliorate this behaviour, we should include the
bound state explicitly as an $O(N)$ singlet field, then amongst the
new set of fixed points in this enlarged space will be one with the
same universal properties as the original $N$ vector model, but with
better behaved derivative expansion properties. Similar considerations
should apply to fixed points with fermions, particularly since the bound state
fields here also correspond to the order parameter (\aka fermion
condensate).\cite{yuri}

The most impressive example so far however,
is provided by the case of $D=2$ dimensions.
It had been conjectured by Zamalodchikov \cite{zam} that there should exist
an infinite series of multicritical points in the two dimensional $Z_2$
(\ie $\phi\leftrightarrow-\phi$) symmetric theory of a single scalar field,
corresponding to the so-called unitary minimal models
in CFT (conformal field theory).
However a verification of this conjecture is in practice well outside the
capabilities of the standard \nonp\ approximation methods: 
the corresponding $\epsilon$ expansions are so badly behaved as
to be useless,\footnote{
\hbox{\it E.g.}{} already with the tricritical point,  $O(\epsilon^2)$ 
underestimates $\eta$ by a factor $\sim1/100$. 
The situation gets factorially worse
as multicriticality is increased.\cite{badE}}
with similar difficulties expected in resummed weak coupling perturbation
theory, while lattice methods suffer from difficulties locating and
accurately computing the multicritical points in these high dimensional
bare coupling constant spaces.\footnote{Indeed to date, only the
lattice computation of the two lowest operator dimensions around the
tricritical point,  
has  been attempted.\cite{asor}} 
All these methods also get rapidly worse with increasing operator dimension.
In constrast, at $O(\partial^2)$, the lowest order at which a fixed point
search through $\eta>0$ can be done, we uncover 
the multicritical points, and {\sl only} these,\footnote{
The search was restricted to real $Z_2$--symmetric
$V$ and $K$, with $K(0)>0$.}
and find an agreement with CFT that {\sl improves} 
with increasing multicriticality
and dimension.\cite{twod} The results are displayed in table 2, for the
first 10 (multi)critical points and up to the first 10 operators. 
\begin{table} [ht]
\centerline{
\vbox{\offinterlineskip\hrule\halign{&\vrule#&\strut\ #\ \hfil\cr
&\hfil $m$&&\hfil$\eta$&&\hfil$\nu$&&\hfil$3^{\rm rd}$&&\hfil$4^{\rm th}$&&
\hfil$5^{\rm th}$&&\hfil$6^{\rm th}$&&\hfil$7^{\rm th}$&&
\hfil$8^{\rm th}$&&\hfil$9^{\rm th}$&&\hfil$10^{\rm th}$&\cr
\noalign{\hrule}
   &\hfil2&&.309&&.863&&.841$^+$&&2.61$^-$&& &&
             && && && && &\cr
     & &&.25&&1&&1$^+$&& && &&
             && && && && &\cr \noalign{\hrule}
     &\hfil3&&.200&&.566&&.234$^+$&&.732$^-$&&1.09$^+$&&
        2.11$^-$&&2.44$^+$&&2.71$^-$&& && &\cr
       & &&.15&&.556&&.2$^+$&&.875$^-$&&1.2$^+$&&
   && && && && &\cr \noalign{\hrule}
     &\hfil4&&.131&&.545&&.166$^+$&&.287$^-$&&.681$^+$&&
     .953$^-$&&1.26$^+$&&2.11$^-$&&2.16$^+$&&2.38$^-$&\cr
      & &&.1&&.536&&.133$^+$&&.25$^-$&&.8$^+$&&
1.05$^-$&&1.33$^+$&& && && &\cr
  \noalign{\hrule}
    &\hfil5&&.0920&&.531&&.117$^+$&&.213$^-$&&.323$^+$&&
     .650$^-$&&.865$^+$&&1.11$^-$&&1.37$^+$&&2.08$^-$&\cr
   & &&.0714&&.525&&.0952$^+$&&.179$^-$&&.286$^+$&&
  .75$^-$&&.952$^+$&&1.18$^-$&&1.43$^+$&& &\cr \noalign{\hrule}
    &\hfil6&&.0679&&.523&&.0868$^+$&&.159$^-$&&.249$^+$&&
     .348$^-$&&.629$^+$&&.806$^-$&&1.01$^+$&&1.22$^-$&\cr
    & &&.0536&&.519&&.0714$^+$&&.134$^-$&&.214$^+$&&
     .313$^-$&&.714$^+$&&.884$^-$&&1.07$^+$&&1.28$^-$&\cr \noalign{\hrule}
    &\hfil7&&.0521&&.517&&.0667$^+$&&.123$^-$&&.193$^+$&&
     .277$^-$&&.368$^+$&&.613$^-$&&.764$^+$&&.933$^-$&\cr
    & &&.0417&&.514&&.0556$^+$&&.104$^-$&&.167$^+$&&
     .243$^-$&&.333$^+$&&.688$^-$&&.833$^+$&&.993$^-$&\cr \noalign{\hrule}
   &\hfil8&&.0412&&.514&&.0529$^+$&&.0972$^-$&&.154$^+$&&
     .221$^-$&&.299$^+$&&.383$^-$&&.601$^+$&&.733$^-$&\cr
   & &&.0333&&.511&&.0444$^+$&&.0833$^-$&&.133$^+$&&
     .194$^-$&&.267$^+$&&.350$^-$&&.667$^+$&&.794$^-$&\cr \noalign{\hrule}
   &\hfil9&&.0334&&.511&&.0429$^+$&&.0790$^-$&&.125$^+$&&
     .180$^-$&&.245$^+$&&.317$^-$&&.395$^+$&&.592$^-$&\cr
   & &&.0273&&.509&&.0364$^+$&&.0682$^-$&&.109$^+$&&
     .159$^-$&&.218$^+$&&.286$^-$&&.364$^+$&&.650$^-$&\cr \noalign{\hrule}
   &10&&.0277&&.509&&.0355$^+$&&.0654$^-$&&.103$^+$&&
     .150$^-$&&.203$^+$&&.265$^-$&&.332$^+$&&.405$^-$&\cr
  & &&.0227&&.508&&.0303$^+$&&.0568$^-$&&.0909$^+$&&
     .133$^-$&&.182$^+$&&.239$^-$&&.303$^+$&&.375$^-$&\cr \noalign{\hrule}
  &11&&.0233&&.508&&.0299$^+$&&.0550$^-$&&.0870$^+$&&
     .126$^-$&&.172$^+$&&.224$^-$&&.282$^+$&&.345$^-$&\cr
  & &&.0192&&.506&&.0256$^+$&&.0481$^-$&&.0769$^+$&&
     .112$^-$&&.154$^+$&&.202$^-$&&.256$^+$&&.317$^-$&\cr
}\hrule}}
{\caption[]
{Critical exponents, and dimensions and $Z_2$ 
parities of the 10 lowest dimension
operators, for the first 10 multicritical points, $m=2\cdots11$.
Thus, $m=2$ is Ising, $m=3$ is tricritical Ising, 
$m=4$ is tetracritical (\aka three-state Potts model), \etc
For each $m$, the 
$O(\partial^2)$ answer is in the first row 
and the associated exact CFT result is in the second row.
}}
\end{table}
We see that
there is a remarkable agreement between these thus lowest order results
and CFT, spanning over two orders of magnitude. The worst 
determined number is $\eta$ for the tricritical point, which is
only accurate to 33\%, but this gradually improves as $m$ increases.
(At $m=11$, $\eta$ is off by 21\%.)
 $\nu$ is worst determined at $m=3$ (13\%)
after which all are determined to error less than 2\% and decreasing with
increasing $m$. Indeed, the best determined number is
$\nu$ for $m=11$, which is accurate to 0.2\%.
The worst determined operator dimension is the $3^{\rm rd}$
at $m=5$
(25\%), after which errors decrease with increasing $m$ and/or increasing 
dimension, so that all the rest have errors in the range 9\% -- 22\%. 
Note that the low dimension operators at high order of multicriticality 
correspond to renormalization group
eigenvalues which agree with CFT to better than 3 significant
figures. Fig.2 shows the fixed point solutions 
for the first three critical points.

We also found 
some irrelevant operators  which cannot be matched
to operators in the CFT minimal models. This is the reason for the blank
spaces in some corresponding CFT parts of the table.
Examples of such
operators have also been found (in the correction to scaling
at the Ising critical point) by $\epsilon$ expansion and fixed dimension
resummed perturbation theory,\cite{zinn} 
and argued for in exact treatments.\cite{schroer}
Further work is required to understand their true significance.

\begin{figure}[ht]  
\epsfxsize=0.9\hsize
\epsfbox{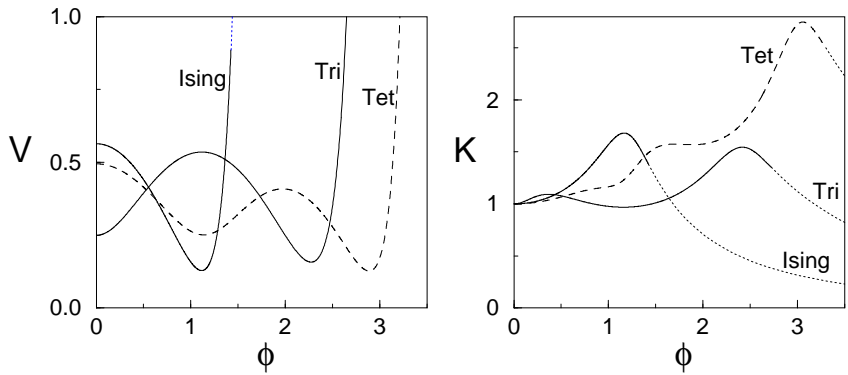} 
\vskip -2cm
\caption{The potential $V(\phi)$, and kinetic term coefficient
function $K(\phi)$, for the two dimensional
Ising, tricritical Ising, and tetracritical Ising
model fixed points.}
\end{figure}

Finally, what are the problems with this approach? The treatment of
gauge theory present special problems: the formulation of a flow equation which
properly treats the quantum aspects of gauge invariance \nonp ly,
and secondly, the construction of reliable approximations.
The solution to the former must proceed in one of two ways:
by allowing the cutoff
to break the gauge invariance and then attempting to recover it
as the cutoff is removed (via broken Ward identities),\cite{ell}
or by generalizing the
flow equations so that gauge invariance is not broken by the cutoff.\cite{ui}
While the first method can be shown to work to all orders in perturbation
theory,\cite{bon} it seems hopeless 
\nonp ly.\cite{qap} It
is surely through an exact preservation of the quantum
gauge invariance that real progress will be made; 
a useable generalization of the pure $U(1)$ case \cite{ui} may well be possible.
Note however, that for the interesting cases (non-Abelian and greater than two
dimensions), derivative expansions {\it per se} 
are anyway impractical due to the large number
of independent gauge invariant combinations that can be formed even at
lowest non-trivial order.\cite{ui} Instead, much more appropriate forms of
approximation deserve study in this case, such as large $N$ methods.

Returning to non-gauge theories,  apart from the
practical problem that higher orders in the derivative expansion get
rapidly more complicated, one  problem that this
method shares with all other approximations to the 
renormalization group \cite{rsRG}\footnote{and analogously perturbation theory
to a certain extent, through scheme dependence.}  
is an unphysical dependence on the choice of cutoff function. 
This dependence is not really a problem however if instead it is used
to estimate a rough  lower bound
 on the error of the approximation, 
and thus test the numerical reliability.\cite{ball}
For example, we can
compare the $O(\partial^0)$ $N=1$ results in table 1, to the corresponding
results for sharp cutoff, $\nu=.70$ and $\omega=.60$.\cite{hashas,deriv}
Another problem that has already been mentioned is that generically derivative
expansions depend also
on one unphysical parameter in the fixed point solutions, due to loss
of reparametrization invariance. In a similar way this is not a problem
(if one knows about it!), but rather can be used to test numerical
reliability in such approximations.\cite{deriv,gol} The real question that needs
to be answered ultimately, is whether the derivative expansion 
exists to all orders and converges,
because of course if it does, these problems have to lessen and eventually
disappear as the expansion is pushed to higher orders. This question is
hard to answer in generality, but
rather straightforward to analyse perturbatively
for a specific theory. In table 3, I show which results converge for the
$\beta$ function of the $O(N)$ invariant scalar field theory in $D=4$
dimensions, computed to
two loops with different forms of derivative expansion.\cite{prep} Also
summarised are which expansions preserve reparametrization invariance.

\begin{table} [ht]
\renewcommand{\arraystretch}{1.5}
\hspace*{\fill}
\begin{tabular}{|c||c||c|c|}  \hline
Variant. & Repar. Inv. &\multicolumn{2}{c|}{Convergence.}
\\ \hline & & one loop & two loops
\\ \hline Wilson/Polchinski & X & X & X
\\ \hline Legendre Power & $\surd$ & $\surd$ & X
\\ \hline Legendre Faster & X & $\surd$ & $\surd$
\\ \hline Legendre Sharp  & $\surd$ & $\surd$ & $\surd$
\\ \hline
\end{tabular}
\hspace*{\fill}
\renewcommand{\arraystretch}{1}
{\caption[Properties of derivative expansions]
{Properties of derivative expansions for different forms of flow equation
and cutoff: Wilson or Polchinski flow equation
with any smooth cutoff, Legendre 
flow equation with the
Power law cutoff described in the text, Legendre flow equation
with some Faster falling
cutoffs, and momentum scale expansion of the Legendre flow equation with
Sharp cutoff.
}}
\end{table}

A direct derivative expansion of the Wilson effective action
does not converge already at one loop: it is again a
result of expansion of the propagators inside the effective action 
(mentioned earlier). On the other hand, the Legendre flow equations
give the exact answer to the one loop $\beta$ function already at 
$O(\partial^0)$.\cite{erg} Only the last two methods however converge
at two loops. In particular, the sharp cutoff case converges very fast at
two loops,\cite{truncm} and since it also preserves a simple 
reparametrization invariance, further work to overcome the practical
difficulties in its implementation \cite{truncm} certainly seems called for.
Although  convergence
even to all orders in perturbation theory, is no guarantee of convergence
\nonp ly (truncations in increasing powers of the field trivially must converge
at any fixed order of perturbation theory, but as we have seen, fail to
do so \nonp ly), the natural conjecture is that these last two methods do
converge \nonp ly. The fact that the Legendre $O(\partial^0)$ approximation
is exact at $N=\infty$ \cite{wegho,oN} lends
 further support to this conjecture.\footnote{\hbox{\it E.g.}{} the
same is not true of truncations in powers of the field, to any order.}

Of course, 
negative answers to convergence do not exclude the 
first two methods from being good model
approximations at low orders.
Although I have concentrated on the Legendre flow equation with power law
cutoff, derivative expansions of
the Wilson / Polchinski equation are distinguished
by their relative simplicity. Perhaps, by
generalising the derivative expansion, one
 can preserve this simplicity while also 
preserving more of the structure of the exact renormalization group.

We remind that a large number of references to other work on 
the lowest order sharp cutoff
approximation have been collected.\cite{truncm} We collect here
 corresponding
smooth cutoff versions
not so far mentioned,\cite{kogwil,smoo} and similarly
attempts to go beyond leading order in the derivative 
expansion.\cite{beyo} There are also a number of examples 
that entertain the idea
of derivative expansion but in practice make further
 truncations.

In conclusion, the derivative (momentum scale) expansion methods --
where no other approximation is made -- are
potentially very powerful, particularly in genuinely \nonp\ settings
where all other methods fail. 
In contrast to more severe truncations, all these variants are 
robust,  in the sense that no spurious solutions have been found, while
especially the Legendre flow equation with power law
cutoff yields very satisfactory numerical
accuracy at low orders. The full potential of these methods is by no means
yet realised, and much more theoretical progress on their 
properties is possible and expected.

\section*{Acknowledgments}
It is a pleasure to thank the organizers of RG96 for the invitation to speak
at this very stimulating conference, several participants -- particularly Yuri 
Kubyshin for helpful conversations, and the SERC/PPARC for financial 
support through an Advanced Fellowship. 

\end{document}